\documentclass
[prc,twocolumn,showpacs,preprintnumbers,amsmath,amssymb,superscriptaddress,floatfix,nofootinbib]{revtex4}
\usepackage{graphicx}
\usepackage{amsmath}
\usepackage{amsfonts}
\usepackage{amssymb}%

  \def\CL{{\cal L}}

\begin{document}

\title{$K^- p \rightarrow \eta \Lambda$ reaction in an effective Lagrangian model}

\author{Bo-Chao Liu} \email{liubc@xjtu.edu.cn} \affiliation{Department of Applied Physics, Xi'an
Jiaotong University, Xi'an, Shanxi 710049, China}

\author{Ju-Jun Xie} \email{xiejujun@mail.ihep.ac.cn}
\affiliation{Department of Physics, Zhengzhou University, Zhengzhou,
Henan 450001, China} \affiliation{Instituto de F\'\i sica
Corpuscular (IFIC), Centro Mixto CSIC-Universidad de Valencia,
Institutos de Investigaci\'on de Paterna, Aptd. 22085, E-46071
Valencia, Spain}

\begin{abstract}

We report on a theoretical study of the $K^- p \to \eta \Lambda$
reaction near threshold by using an effective Lagrangian approach.
The role of $s-$channel $\Lambda(1670)$, $t-$channel $K^*$ and
$u-$channel proton pole diagrams are considered. We show that the
total cross sections data are well reproduced. However, only
including the $s-$wave $\Lambda(1670)$ state and the background
contribution from $t-$ and $u-$channel are not enough to describe
the bowl structures in the angular distribution of $K^- p \to \eta
\Lambda$ reaction, which indicates that there should be higher
partial waves contributing to this reaction in some energy region.
Indeed, if we considered the contributions from a $D_{03}$
resonance, we can describe the bowl structures, however, a rather
small width ($\sim 2$ MeV) of this resonance is needed.

\end{abstract}
\maketitle

The $K^-$ induced reactions are important tool to gain a deeper
understanding of the $\bar{K}N$ interactions and also of the nature
of the hyperon resonance. The reaction $K^- p \to \eta \Lambda$ is
of particular interest in the hyperon resonances since there are no
isospin-1 hyperons contributing here and it gives us a rather clear
channel to study the $\Lambda$ resonances. Ten years ago, the
differential and total cross sections of the $K^- p \to \eta
\Lambda$ reaction have been measured, with much higher precision
than previous measurements, by the Crystal Ball
Collaboration~\cite{data}. These new data are obtained with beam
momentum of $K^-$ from threshold to 770 MeV/c, corresponding to
invariant mass $\sqrt{s}=1.664-1.685$ GeV.

Current knowledge of $\Lambda$ resonances are mainly known from the
analysis of $\bar K N$ reactions in the 1970s, and large
uncertainties exist because of poor statistics of data and limited
knowledge of background contributions~\cite{pdg2010,gaopz2011}.
Besides, the nature of some $\Lambda$ states are still
controversial. Based on the available new data with much higher
precision, the authors of Ref.~\cite{data} come to the conclusion
that $\Lambda(1670)$ should be a three-quark state, while on the
contrary the authors of Refs.~\cite{oset1,oset2} argue that
$\Lambda(1670)$ is a dynamically generated state. On the other hand,
the traditional three-quark features of $\Lambda(1670)$ are shown in
Ref.~\cite{zhongprc79} from a studying $K^-p \to \pi^0 \Sigma^0$
reaction at low energies by using a chiral quark model. It is clear
that some further and detailed studies, both on theoretical and
experimental sides, are still necessary.

Since the $\Lambda(1670)$ has large coupling to the $\bar K N$ and
$\eta \Lambda$ channels, it is expected that $\Lambda^*$ should
dominate this reaction near threshold. In the present work, we
reanalyze the $K^- p \to \eta \Lambda$ reaction near threshold
within the effective Lagrangian method. In addition to the main
contribution from $\Lambda(1670)$ state, the "background"
contributions from the $t-$channel $K^*$ exchange and the
$u-$channel proton exchange are also studied.

\begin{figure}[htbp]
\begin{center}
\includegraphics[scale=0.4]{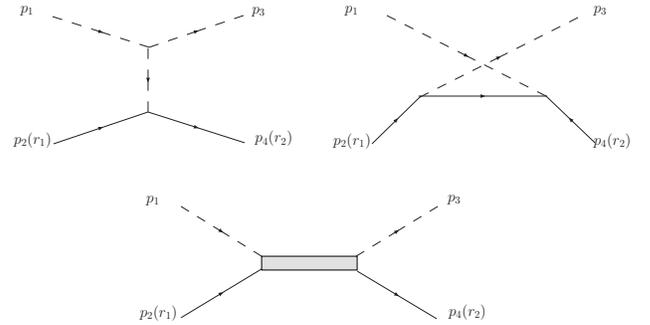}
\caption{Model for the reaction $K^- p \to \eta \Lambda$. In these
diagrams, we show the definition of the kinematical ($p_1, p_2, p_3,
p_4$) and polarization variables $r_1, r_2$ that we use in our
calculation.} \label{feynfig}
\end{center}
\end{figure}

The basic Feynman diagrams are shown in Fig.~\ref{feynfig}. These
include $t-$channel $K^*$ exchange, $u-$channel proton exchange, and
the $s-$channel $\Lambda(1670) (\equiv \Lambda^*)$ terms. To compute
the contributions of these terms, we use the interaction Lagrangian
densities of Refs.~\cite{wufq,xie,Mosel,feuster}:
\begin{eqnarray}
\CL_{K^* K \eta} &=&
g_{K^*K\eta}(\eta\partial^\mu{K^-}-K^-\partial^\mu{\eta})K^{*-}_\mu
\label{ksnl} \\
{\cal L}_{K^*N\Lambda}&=& g_{K^*N\Lambda}\bar\Lambda\big(\gamma_\mu
-{\kappa \over 2M_N}\sigma_{\mu\nu}\partial^\nu
\big)K^{*\mu}N \nonumber \\ && + \mathrm{H.c.} \,, \\
{\cal L}_{\eta NN}&=&g_{\eta NN}{\bar N}\gamma_5 N\eta, \\
{\cal L}_{KN\Lambda}&=&g_{KN\Lambda}\bar{N}\gamma_5\Lambda
K+\mathrm{H.c.}, \\
{\cal L}_{\Lambda^*{\bar K}N}&=&g_{\Lambda^*
{\bar K}N} \bar\Lambda^* \bar{K} N+\mathrm{H.c.}, \\
{\cal L}_{\Lambda^* \Lambda\eta}&=&g_{\Lambda^* \Lambda\eta}
\bar{\Lambda}^* \eta \Lambda +\mathrm{H.c.}.
\end{eqnarray}
where we take $\kappa =2.43$ that determined by the Nijmegen
potential~\cite{stoks99} and has been used in Ref.~\cite{oh06}.
Other coupling constants will be discussed below.

With the effective Lagrangian densities given above, we can easily
construct the invariant scattering amplitudes:
\begin{eqnarray}
{\cal M}_i=\bar u_{r_2}(p_4)~{\cal A}_i~u_{r_1}(p_2),
\end{eqnarray}
where $i$ denotes the $i$th channel that contributes to the total
amplitude, and $\bar u_{r_2}(p_4)$ and $u_{r_1}(p_2)$ are the
spinors of $\Lambda$ and proton, respectively. The reduced ${\cal
A}_i$ read
\begin{eqnarray}
{\cal A}_s &=& g_{\Lambda^* {\bar K}N}g_{\Lambda^*
\Lambda\eta} \frac{{\not \! p_1}+{\not \! p_2}+M_{\Lambda^*}}{ s - M_{\Lambda^*}^2+iM_{\Lambda^*}\Gamma_{\Lambda^*}} , \\
{\cal A}_t &=&  i  \frac{g_{K^*K\eta} g_{K^*\Lambda
N}}{q^2-m^2_{K^*}} ( {\not \! p_1}+{\not \! p_3}-
\frac{m^2_{K}-m^2_{\eta}}{m^2_{K^*}} {\not \! q} \nonumber \\
&& -\frac{\kappa}{m_N}(p_1\cdot p_3-{\not \! p_1}{\not \! p_3}) ) , \\
{\cal A}_u &=& - ~g_{K\Lambda N} g_{\eta NN} \frac{{\not \!
p_2}-{\not \! p_3}-m_N}{ u - m^2_N}.
\end{eqnarray}
where $q$ is the momentum of exchanging meson $K^*$ in the
$t-$channel. The width of $K^*$ is not taken into account since
$K^*$ is in the $t-$channel. The subindices $s, t,$ and $u$ stand
for the $s-$ channel $\Lambda^*$ exchange, $t-$channel $K^*$
exchange, and $u-$channel proton pole terms. As we can see, in the
tree-level approximation, only the products like $g_{\Lambda^*{\bar
K} N} g_{\Lambda^* \Lambda \eta}$, enter in the invariant
amplitudes. They are determined with the use of MINUIT, by fitting
to the experimental data~\cite{data}, including the total and
differential cross sections. Besides, $M_{\Lambda^*}$ and
$\Gamma_{\Lambda^*}$ are the mass and total decay width of the
$\Lambda^*$ resonance, which are free parameters in the present work
and will be also fitted to the experimental data.

Because we are not dealing with point-like particles, we ought to
introduce the compositeness of the hadrons. This is usually achieved
by including form factors in the amplitudes. In the present work, we
adopt the following form factors~\cite{wufq,Mosel,feuster}
\begin{equation}\label{FB}
F(q_{ex}^2,M_{ex})={\Lambda^4\over
\Lambda^4+(q_{ex}^2-M_{ex}^2)^2}\, ,
\end{equation}
for $s-$ and $u-$channel, and
\begin{equation}\label{FB2}
F(q_{ex}^2,M_{ex})=\left
(\frac{\Lambda^2-M_{ex}^2}{\Lambda^2-q_{ex}^2}\right )^2\, ,
\end{equation}
for $t-$channel, where the $q_{ex}$ and $M_{ex}$ are the 4-momenta
and the mass of the exchanged hadron, respectively. For the cutoff
parameters, we take $\Lambda=2.0$ GeV for $s-$channel, $\Lambda=1.5$
GeV for $t-$ and $u-$channel.

The differential cross section for $K^-p \to \eta \Lambda$ at center
of mass (c.m.) frame can be expressed as
\begin{equation}
{d\sigma \over d{\rm cos}\theta_{\rm c.m.}}={1\over 32\pi s}{
|\vec{p_3}^{\text{c.m.}}| \over |\vec{p_1}^{\text{c.m.}}|} \left (
{1\over 2}\sum_{r_1,r_2}{|\cal M|}^2 \right ), \label{eq:dcs}
\end{equation}
where $\theta_{\rm c.m.}$ denotes the angle of the outgoing $\eta$
relative to beam direction in the $\rm c.m.$ frame, and
$s=(p_1+p_2)^2$, is the invariant mass square of the system.

In Eq.~(\ref{eq:dcs}), the total invariant scattering amplitude
$\cal M$ is given by,
\begin{equation}
{\cal M}={\cal M}_s + e^{i\theta_1} {\cal M}_t+ e^{i\theta_2} {\cal
M}_u \, .
\end{equation}

In the phenomenological Lagrangian approaches, the relative phases
between amplitudes from different diagrams are not fixed, so we
introduce two relative phases $\theta_1$ and $\theta_2$ between the
background and the $\Lambda^*$ contributions as free parameters,
which will be determined by fitting to the experimental data.

We perform seven-parameter ($M_{\Lambda^*}$, $\Gamma_{\Lambda^*}$,
$g_{\Lambda^* {\bar K}N}g_{\Lambda^* \Lambda\eta}$,
$g_{K^*N\Lambda}g_{K^*K\eta}$, $g_{KN\Lambda}g_{\eta NN}$,
$\theta_1$, and $\theta_2$) $\chi^2$ fit to the total and
differential cross section data taken from Ref.~\cite{data}. There
are a total $155$ data points.

The fitted parameters for $\Lambda(1670)$ are shown in
Table.~\ref{tab1} and other fitted results are:
$g_{K^*N\Lambda}g_{K^*K\eta}= 14.8 \pm 1.7$, $g_{KN\Lambda}g_{\eta
NN} = -5.6 \pm 0.9$, $\theta_1=2.9 \pm 0.2$, and $\theta_2=2.9 \pm
0.3$. The resultant $\chi^2/dof$ is $1.3$.

\begin{table}[htbp]
\caption{Adjusted parameters for $\Lambda(1670)$ resonance. PDG
estimates are also listed for comparison. }
\begin{tabular}{cccc}
\hline\hline    & Mass(MeV) & $\Gamma_\mathrm{tot}$(MeV) & $|g_{\Lambda^*{\bar K}N} g_{\Lambda^*\eta \Lambda}|$\\
\hline
This calculation & $1671.5\pm 0.2$ & $23.3\pm 0.2$ & $0.28 \pm 0.03$ \\
PDG &$1660 \sim 1680$ &$25 \sim 50$ & $0.31 \pm 0.15$
\\\hline \hline
\end{tabular}
\label{tab1}
\end{table}

On the other hand, the coupling constants of $g_{\Lambda^* {\bar K}
N}$ and $g_{\Lambda^* \eta \Lambda}$ can be also evaluated from the
$\Lambda(1670)$ to ${\bar K} N$ and $\eta \Lambda$ partial decay
widths:
\begin{eqnarray}
\Gamma_{\Lambda^* \to {\bar K}N} &=& \frac{g^2_{\Lambda^*{\bar
K}N}}{2\pi} (E_N+m_N)\frac{|\vec{p}_N|}{M_{\Lambda^*}}, \\
\Gamma_{\Lambda^* \to \eta \Lambda} &=& \frac{g^2_{\Lambda^*\eta
\Lambda}}{4\pi}
(E_{\Lambda}+m_{\Lambda})\frac{|\vec{p}_{\Lambda}|}{M_{\Lambda^*}},
\end{eqnarray}
where
\begin{eqnarray}
E_{N/\Lambda} &=& \frac{M^2_{\Lambda^*}+m^2_{N/\Lambda}-m_{{\bar
K}/\eta}}{2M_{\Lambda^*}},\\
|\vec{p}_{N/\Lambda}| &=& \sqrt{E^2_{N/\Lambda}-m^2_{N/\Lambda}} \,
.
\end{eqnarray}

With the value of total decay width $\Gamma_{\Lambda^*} = 35 \pm 15$
MeV, a value of $0.25 \pm 0.05$ for the $\Lambda^* \to {\bar K}N$
branching ratio, and a value of $0.175 \pm 0.075$ $\Lambda^* \to
\eta \Lambda$ branching ratio, quoted in the Particle Data Group
(PDG) book~\cite{pdg2010}, we can get $|g_{\Lambda^* {\bar
K}N}g_{\Lambda^* \eta \Lambda}| = 0.31 \pm 0.15$, which was also
shown in Table. I for comparison. The error $\pm 0.15$ is came from
that the errors of the $\Lambda^* \to {\bar K}N$ and $\Lambda^* \to
\eta \Lambda$ partial decay widths.

As we can see in Table. I, the fitted parameters for the
$\Lambda(1670)$ resonance agree well with that of the PDG
estimation.

During the best fit, we adjusted the product of the coupling
constants to experimental data. If we take $g_{K^*K\eta}=1.6$ that
was obtained from the $SU(3)$ prediction~\cite{wufq}, then we can
get $ |g_{K^*N\Lambda}| = 9.3 \pm 1.0$ which roughly agrees with the
value, $ |g_{K^*N\Lambda}| = 6.1$, which was obtained from the
$SU(3)$ flavor symmetry in Ref.~\cite{stoks99}. Since the value of
$g_{\eta NN}$ is extremely uncertain and if we adopt it as $2.24$
that was used in Ref.~\cite{xie}, then we get $|g_{KN\Lambda}| = 2.5
\pm 0.5$ which is much different with the $SU(3)$ prediction value
$13.3$~\cite{ohprt06,oh08}. However, as we mentioned above, the
uncertainty of $g_{\eta NN}$ is very
large~\cite{nneta1,nneta2,nneta3,nneta4,nneta5,nneta6}, so the adjusted coupling
constant $g_{KN\Lambda}$, in the present work, may be still within
the $SU(3)$ prediction.

\begin{figure}[htbp]
\begin{center}
\includegraphics[scale=0.7]{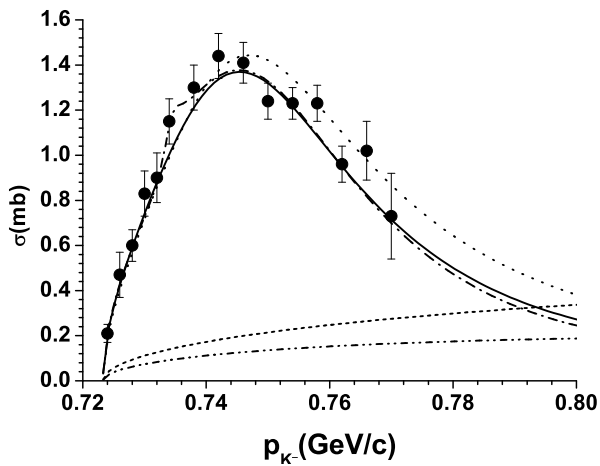} \vspace{-0.2cm}
\caption{$K^- p \to \eta \Lambda$ total cross sections compared with
the data~\cite{data}. Results have been obtained from the best
$\chi^2$ fit. The solid line represents the full results, while the
contribution from $\Lambda(1670)$, $t-$, and $u-$channel diagrams
are shown by the dotted, dashed and dot-dot-dashed lines,
respectively. The dot-dashed represents the best results for the total
cross sections after including the $D_{03}$ state.} \label{tcs}
\end{center}
\end{figure}

Our best fits to the experimental data of the total cross sections
are shown in Fig.~\ref{tcs}, comparing with the data. The solid line
represents the full results, while the contribution from
$\Lambda(1670)$, $t-$, and $u-$channel diagrams are shown by the
dotted, dashed and dot-dot-dashed lines, respectively. From
Fig.~\ref{tcs}, one can see that we can describe the data of total
cross sections quite well and the $\Lambda(1670)$ gives the dominant
contribution, while the $t-$ and $u-$channel diagrams give the minor
but sizeable contribution.

\begin{figure*}[htbp]
\begin{center}
\includegraphics[scale=0.85]{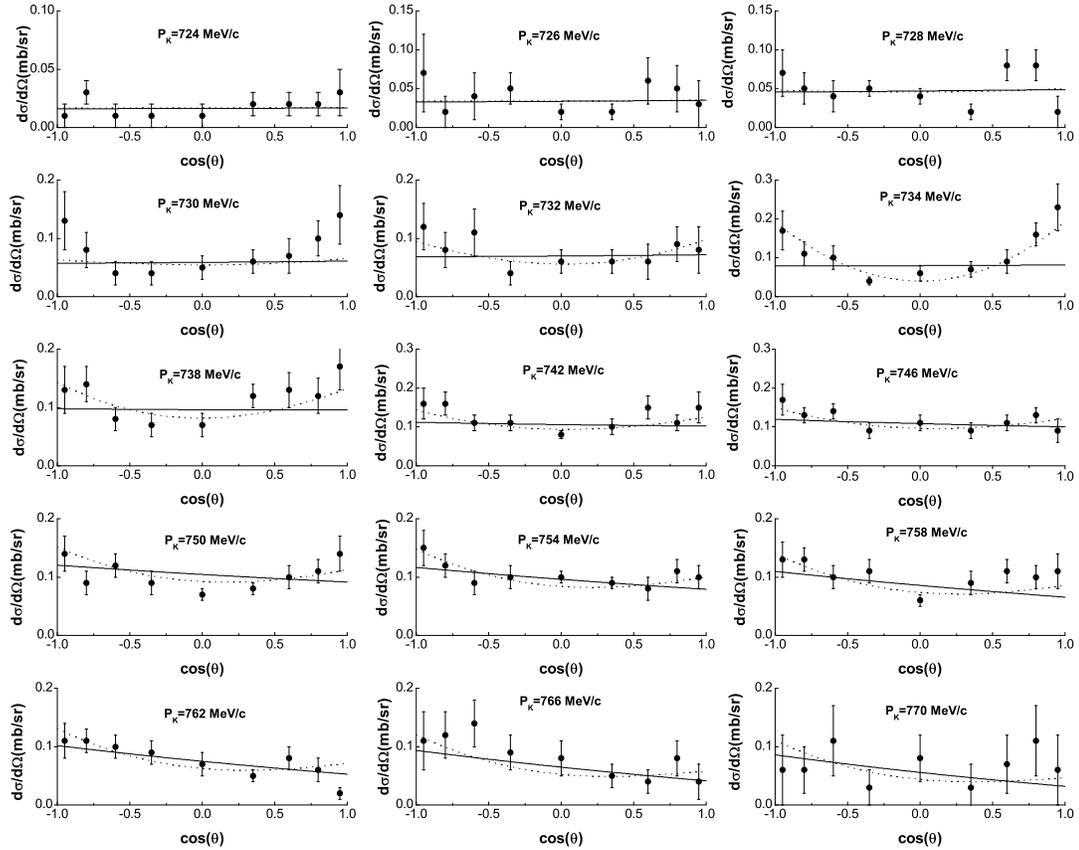} \vspace{-1.5cm}
\caption{The best fitting results for differential cross sections.
The solid lines represent the results by considering only
$\Lambda(1670)$ and background contributions, while the dashed lines
represents the result by including also a narrow $D_{03}$
resonance.} \label{dcs}
\end{center}
\end{figure*}

The results of the best fit for the differential cross sections are
shown with the solid line in Fig.~\ref{dcs}. From there we can see
that the deviations between our theoretical results and experimental
data are evident especially for the angular distribution at $p_{K^-}
= 730, 732, 734, 738, 742$ MeV, where bowl-shaped structures in
angular dependence appear. It also should be noted that with
including the background contribution from the $t-$channel $K^*$
exchange and $u-$channel proton exchange, the backward enhancement
in the angular distribution for $p_{K^-}$ from $750$ to $770$ MeV
are reproduced.

In order to obtain a better description of the differential cross
section data, especially at some energy points, some other
resonances that may contribute to this reaction should also be
considered. For the bowl structures in differential cross sections,
one possible explanation is that there might be $d-$wave
contributions from the $s-$channel with the excitation of $D_{03}$
resonance. For checking this, we performed another best fit: in
addition to the contributions which were already considered in the
previous fit, the contribution from the $D_{03}$ state in the
$s-$channel process are also included. The new best fitting gives
$\chi^2/dof=0.9$ and we get a satisfied description for both total
cross sections and differential cross sections. The new results for
the total cross sections are similar with the previous results
except for a small bump around $p_{K^-}=736$MeV(see the dot-dashed
line in Fig. \ref{tcs}).
 The corresponding results for differential
cross sections are shown with dotted line in Fig.~\ref{dcs}, where
the bowl structures are well reproduced.

The fitted parameters for $D_{03}$ resonance are mass $M=1668.5 \pm
0.5$ MeV and total decay width $\Gamma=1.5 \pm 0.5$ MeV. The mass of
$D_{03}$ is close to the PDG estimate for $\Lambda(1690)$
($M_{\Lambda(1690)}= 1690 \pm 5$ MeV), while the width is too small
compared to the PDG estimate ($\Gamma_{\Lambda(1690)} = 60 \pm 10$
MeV). The width obtained from the best fit is narrow because the
bowl structures in the differential cross sections are shown up in a
narrow ($\pm 3$ MeV)~\footnote{This is evaluated from the $K^- p$
invariant mass changed, with the range $730-742$ MeV of $p_{K^-}$,
by using the relation $s = (p_1+p_2)^2 = m^2_{K^-} + m^2_p +
2m_p\sqrt{m^2_{K^-}+p^2_{K^-}}$.} energy window.

One might think that releasing the limit of the cutoff values for
the form factors and inclusion of more $\Lambda$ resonances (such as
$\Lambda(1600)$) might improve the situation that the width of the
$D_{03}$ state is too narrow. We have explored such possibility, but
we have found tiny changes. The new best fitting still favor a
$D_{03}$ resonance with very small width and the corresponding
values for the parameters of $D_{03}$ resonance are close to the
values that were obtained above.

In summary, we have studied the $K^- p \to \eta \Lambda$ reaction
near threshold by using an effective Lagrangian approach. The role
of the $s-$channel $\Lambda(1670)$, $t-$channel $K^*$ and
$u-$channel proton pole diagrams are considered. The total cross
section are well reproduced. Our results show that $\Lambda(1670)$
gives the dominant contribution, while the $t-$ and $u-$channel
diagrams give the minor but sizeable contribution, especially for
the backward enhancement in the angular distribution for $p_{K^-}$
from $750$ to $770$ MeV.

However, including $\Lambda(1670)$ resonance in the $s-$channel as
well as the background contributions is not enough to describe the
bowl structures in the angle distributions at some beam momentum
points. A general opinion is that these bowl structures in angular
distribution can be understood by further including the contribution
from $\Lambda(1690)D_{03}$. Indeed, our calculations show that with
considering the $D_{03}$ resonance, we can describe the bowl
structures, but a rather small width of this resonance is needed.
This means that the experimental data can not be understood by
considering the conventional $\Lambda(1690)$. On the other hand, the
current experimental data still have systematic uncertainties
especially when we look at the angular distribution data obtained
from two different ways of identifying the final $\eta$ meson(see
Fig. 20 of Ref.~\cite{data}), so the present results give a signal
for the needs of further studies in this reaction.

\section*{Acknowledgments}

We would like to thank Xu Cao for useful discussions. This work is
partly supported by the National Natural Science Foundation of China
under grants 10905046 and 11105126.

\end{document}